\documentclass[twocolumn,preprintnumbers,amsmath,amssymb]{revtex4}
\usepackage{amssymb}
\usepackage{latexsym}
\usepackage{epsfig}
\usepackage{color}

\begin{document}
\title{Tsallis Holographic Dark Energy}
\author{M. Tavayef$^{1}$, A. Sheykhi$^{1,2}$\footnote{asheykhi@shirazu.ac.ir},
Kazuharu Bamba$^3$\footnote{bamba@sss.fukushima-u.ac.jp}, H.
Moradpour$^{2}$\footnote{Corresponding author:\\
h.moradpour@riaam.ac.ir}}
\address{$^1$ Physics Department and Biruni Observatory, College of
Sciences, Shiraz University, Shiraz 71454, Iran\\
$^2$ Research Institute for Astronomy and Astrophysics of Maragha
(RIAAM), P.O. Box 55134-441, Maragha, Iran\\
$^3$ Division of Human Support System, Faculty of Symbiotic
Systems Science, Fukushima University, Fukushima 960-1296, Japan}

\begin{abstract}
Employing the modified entropy-area relation suggested by Tsallis
and Cirto \cite{non3}, and the holographic hypothesis, a new
holographic dark energy (HDE) model is proposed. Considering a
flat Friedmann-Robertson-Walker (FRW) universe in which there is
no interaction between the cosmos sectors, the cosmic implications
of the proposed HDE are investigated. Interestingly enough, we
find that the identification of IR-cutoff with the Hubble radius, can
lead to the late time accelerated Universe even in the absence of
interaction between two dark sectors of the Universe. This is in
contrast to the standard HDE model with Hubble cutoff, which does
not imply the accelerated expansion, unless the interaction is
taken into account.
\end{abstract}

\maketitle

\section{Introduction}
The last few decades have seen remarkable progress in our
understanding of the Universe. It is almost a general belief that
about ninety five percent of our Universe are composed of two
mysterious components called Dark Matter (DM) and Dark Energy
(DE). The unknown nature of these two components represents some
fundamental questions and
indicates that there is radically new physics to be discovered. In
particular, DM constitutes about $25$ percent of the total energy
density of the Universe. While the existence of DM is firmly
established by astrophysical observations on a wide range of
scales, the nature of DM is still unknown. DE is an even more mysterious
component of our Universe. DE is responsible for the current
accelerated Universe, and it is
very different from ordinary baryonic matter (DE must have a strong
negative pressure). DE contributes nearly $70$ percent of the
energy density of the present Universe. One possibility is that DE
is modeled by the cosmological constant as introduced by
Einstein. However, the required amplitude is very difficult to
reconcile with our understanding of the quantum properties of the
vacuum. Other possibilities are that DE arises from the evolution
of dynamical fields of an unknown origin or it is due to the
modifications of General Relativity. In order to
distinguish observationally between these hypotheses, a sustained
worldwide effort is ongoing to measure the effective equation of
state of DE and its clustering properties using wide field
cosmological surveys \cite{Rev1,Rev2}.

The origin of the current acceleration of the Universe expansion
is a controversial problem in the modern physics \cite{Rev1,Rev2}.
As we mentioned DE theory and modified gravity are two approaches
to explain the late time accelerated expansion investigated widely in the literatures. For reviews on the DE
and theories of modified gravity to explain the late-time cosmic
acceleration, see~\cite{Nojiri:2010wj,Nojiri:2006ri,Book-Capozziello-Faraoni,Capozziello:2011et,
Bamba:2012cp,Joyce:2014kja,Koyama:2015vza,Bamba:2015uma,Nojiri:2017ncd}
and references therein. HDE hypothesis is also a promising approach for
solving the DE puzzle which has arisen a lot of attentions
\cite{HDE,HDE5,shenwang2005,sheykhiphlet2009,zhang2006,sheykhi2012,setarej2010a,sheykhi2009ph,sheykhi2010physlet,karamikhaled20011,
HDE17,HDE01,HDE1,HDE2,HDE3,RevH,wang,stab}. In addition, the
primary model of HDE, based on the Bekenstein entropy and the
Hubble horizon as its IR cutoff, cannot provide suitable
description for the history of a flat Friedmann-Robertson-Walker
(FRW) universe \cite{HDE01,HDE1,HDE2,HDE3,HDE5}. Physicists try to
solve these failures by considering $i$) other cutoffs, $ii$)
probable interactions between the cosmos sectors, $iii$) various
entropies or even, a combination of the mentioned approaches
\cite{RevH,wang}.

Due to the long-range aspect of gravity and in the shadow of the
unknown nature of spacetime, various generalized entropy
formalisms have been employed to study gravitational and
cosmological phenomena
\cite{non2,non15,non16,non17,non18,non19,non20,non21,non22,non23,
non13,non4,non5,non6,non7,non8,non9,non10,non11,non12,non14,eb,eb1,3,4,5,6,7,8,9,10,11,cite1,cite2}.
The results of these studies indicate that the power-law
distributions of probability (generalized entropy formalisms) have
acceptable agreement with gravity and its related issues.

Attributing various generalized entropies to the horizon of FRW
universe, two new HDE models have recently been proposed
\cite{smm,me}. The backbone of these attempts comes from the fact
that the Bekenstein entropy can be obtained by applying the
Tsallis statistics to the system horizon \cite{5,abe,nn1,nn2}. The
obtained models also show satisfactory stability by themselves
\cite{smm,me}.

Indeed, the Tsallis's definition of entropy \cite{non1} plays a
crucial role in studying the gravitational and cosmological
systems in the framework of the generalized statistical mechanics
\cite{non2,non15,non16,non17,non18,non19,non20,non21,non22,non23,
non3,non13,non4,non5,non6,non7,non8,non9,non10,non11,non12,non14,
eb,eb1,3,4,5,6,7,8,9,10,11,cite1,cite2,smm,me}.
As it has been shown by Tsallis and Cirto \cite{non3}, the
Bekenstein entropy is not the only result of applying the Tsallis
statistics to the system. In general, the Tsallis entropy content
of system is a power-law function of the system's area
\cite{non3}, a result which is confirmed by the quantum gravity
considerations \cite{jalal}.

In the present work, using the general model of the Tsallis's
entropy expression \cite{non3}, and taking the holographic
hypothesis into account, we propose a new HDE model for describing
the late time accelerated Universe in sec. II. We also investigate
the evolution of a flat FRW universe, filled by this new HDE and a
pressureless DM in sec. III, by assuming that there is no mutual
interaction between HDE and DM. The age of the present Universe in our model has also been addressed in sec. IV. The last section is devoted to
conclusions.
\section{The Model}
Let us remind that the definition and derivation of standard holographic
energy density ($\rho_{D} =3c^2m^2_p/L^2$) depends on the
entropy-area relationship $S\sim
 A \sim L^2$ of black holes, where $A=4\pi L^2$ represents
the area of the horizon \cite{HDE}. However, this definition of
HDE can be modified due to the quantum considerations \cite{RevH,wang}. It was shown
by Tsallis and Cirto that the horizon entropy of a black hole may
be modified as \cite{non3}
\begin{eqnarray}\label{ent}
S_{\delta}= \gamma A^{\delta},
\end{eqnarray}
where $\gamma$ is an unknown constant and $\delta$ denotes the
non-additivity parameter \cite{non3}. It is obvious that the
Bekenstein entropy is recovered at the appropriate limit of
$\delta=1$ and $\gamma=1/4G$ (in the unit where $\hbar=k_B=c=1$).
In fact, at this limit, the power-law distribution of probability
becomes useless, and the system is describable by the ordinary
probability distribution of probability \cite{non3}. This relation
is also confirmed by the quantum gravity \cite{jalal}, and leads
to interesting results in the cosmological and holographical setups
\cite{non13,non8,non9,non10,non14}.

Based on the holographic principle which states that the number of
degrees of freedom of a physical system should scale with its
bounding area rather than with its volume \cite{Suss1} and it
should be constrained by an infrared cutoff, Cohen et al.,
proposed a relation between the system entropy ($S$) and the IR
($L$) and UV ($\Lambda$) cutoffs as \cite{HDE}
\begin{eqnarray}
L^{3}\Lambda^{3}\leq\left(S\right)^{3/4},
\end{eqnarray}
which after combining with Eq.~(\ref{ent}) leads to \cite{HDE}
\begin{eqnarray}
\Lambda^{4}\leq\left(\gamma(4\pi)^{\delta}\right)L^{2\delta-4}.
\end{eqnarray}
where $\Lambda^{4}$ denotes the vacuum energy density, the
energy density of DE ($\rho_D$) in the HDE hypothesis
\cite{HDE5,HDE17,smm}. Using the above inequality, we can
propose the Tsallis holographic dark energy density (THDE) as
\begin{eqnarray}
\rho_D=BL^{2\delta-4},
\end{eqnarray}
where $B$ is an unknown parameter \cite{HDE5,HDE17,smm}. Let us
consider a flat FRW universe for which the Hubble horizon, is a
proper candidate for the IR cutoff, and there is no interaction
between the DE candidate and other parts of cosmos. In this manner
($L=H^{-1}$), the energy density and conservation law
corresponding to THDE are obtained as
\begin{eqnarray}\label{rho}
&&\rho_D=BH^{-2\delta+4},\\
&&\dot{\rho}_{D}+3H\rho_{D}(1+\omega_{D})=0,\label{6}
\end{eqnarray}
where $\omega_D={P_D}/{\rho_D}$ and $P_D$ denote the equation of
state parameter and pressure of THDE, respectively.
\section{The universe evolution}
For a flat FRW universe filled by THDE and pressureless DM, the
first Friedmann equation takes the form
\begin{eqnarray}\label{frd}
H^{2}=\frac{1}{3m_{p}^{2}}\left(\rho_{D}+\rho_{m}\right),
\end{eqnarray}
where $\rho_{m}$ is the energy density of pressureless matter.
Defining the dimensionless density parameter as
$\Omega_i={\rho_i}/{\rho_c}$, where $\rho_c=3m_{p}^{2}H^{2}$ is
called the critical energy density \cite{HDE5,HDE17}, we can
easily find that

\begin{eqnarray}\label{3}
\Omega_{D}=\frac{\rho_{D}}{3m_{p}^{2}H^{2}}=\frac{B}{3m_{p}^{2}}H^{-2\delta+2}, \ \
\Omega_{m}=\frac{\rho_{m}}{3m_{p}^{2}H^{2}},
\end{eqnarray}
\noindent helping us in rewriting Eq.~(\ref{frd}) as

\begin{eqnarray}\label{u}
\Omega_{m}+\Omega_{D}=\Omega_{D}(1+u)=1,
\end{eqnarray}
\noindent in which
$u=\frac{\rho_{m}}{\rho_{D}}=\frac{\Omega_{m}}{\Omega_{D}}$. Since
THDE does not interact with other parts of cosmos (DM), the
conservation equation of dust is
\begin{eqnarray}\label{5}
\dot{\rho}_{m}+3H\rho_{m}=0.
\end{eqnarray}
\begin{figure}[ht]
\centering
\includegraphics[scale=0.35]{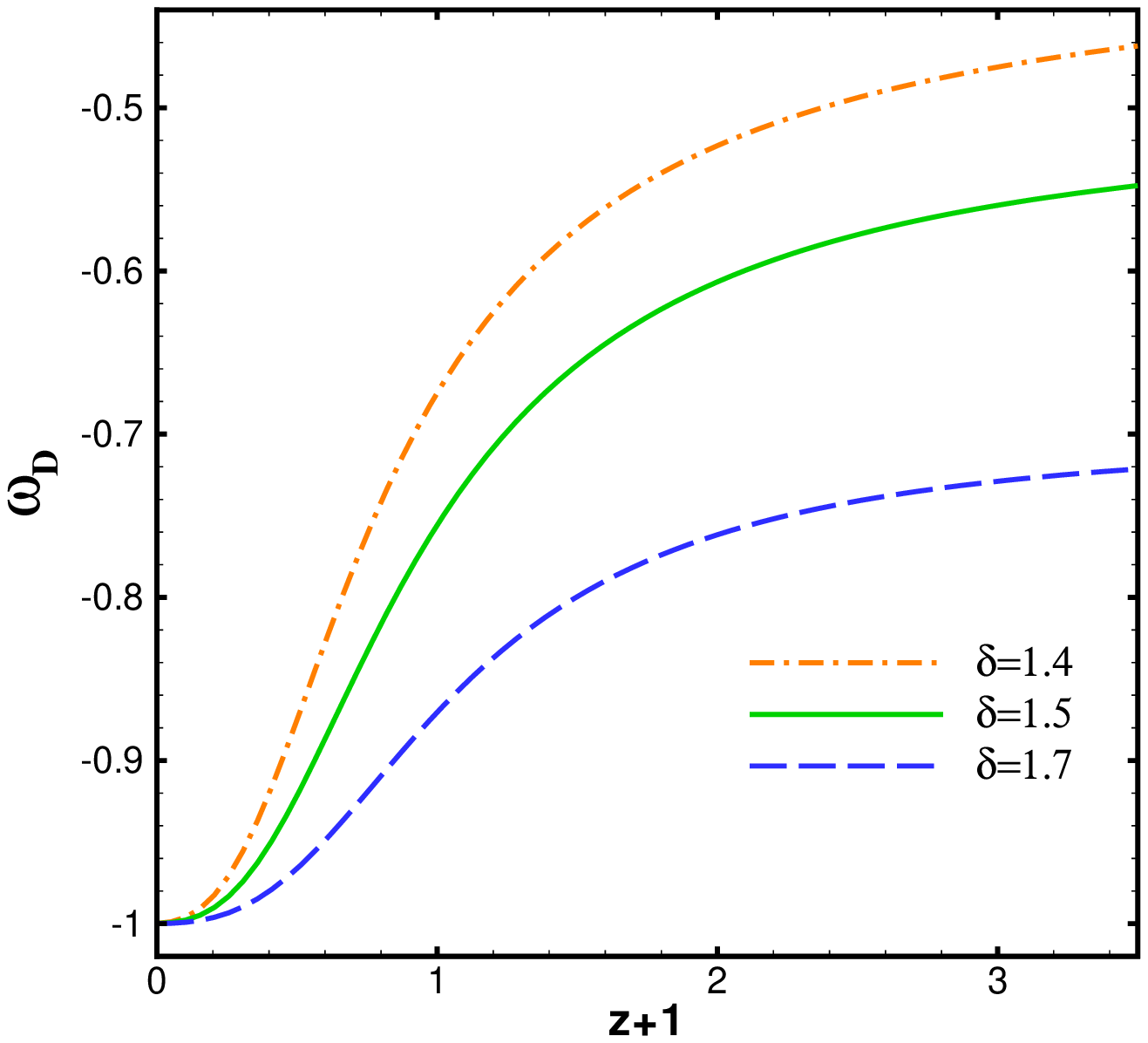}
\includegraphics[scale=0.35]{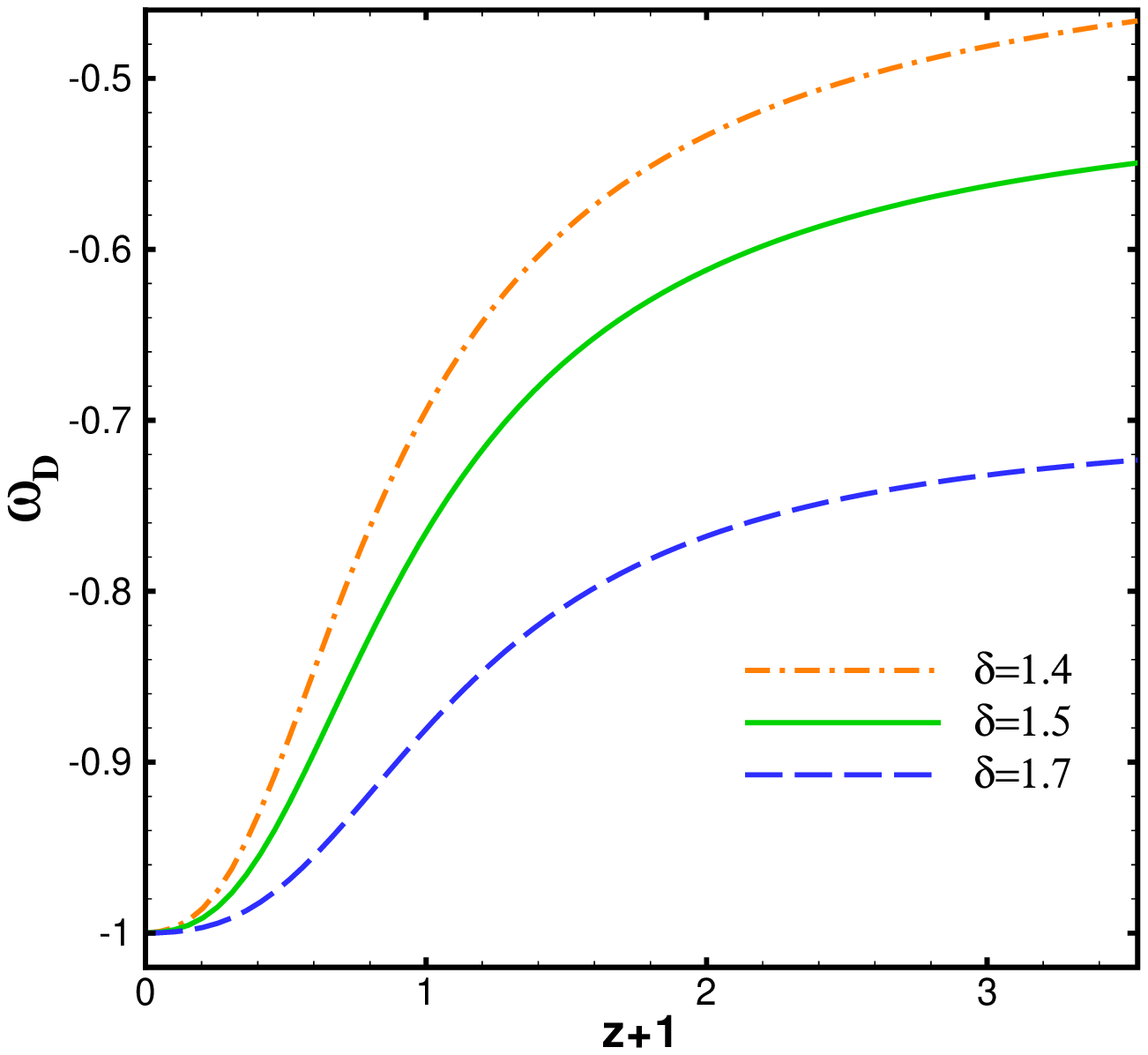}
\caption{$\omega_D$ for some values of $\delta$ as a function of
$z$, whenever $\Omega_D(z=0)=0.70$ (the upper panel) and
$\Omega_D(z=0)=0.73$ (the lower panel). \label{w}}
\end{figure}
Taking the time derivative of Eq.~(\ref{frd}), using
Eqs.~(\ref{5}) and~(\ref{6}), and combining the result with
Eq.~(\ref{3}), one can obtain
\begin{eqnarray}\label{7}
\frac{\dot{H}}{H^{2}}=-\frac{3}{2}(1+\omega_{D}+u)\Omega_{D}.
\end{eqnarray}
On the other hand, inserting Eq.~(\ref{rho}) into Eq.~(\ref{6}),
we find that
\begin{eqnarray}\label{80}
\frac{\dot{H}}{H^{2}}=(1+\omega_{D})\dfrac{3}{2\delta-4},
\end{eqnarray}
compared with Eq.~(\ref{7}) to reach at
\begin{eqnarray}
\omega_{D}=\dfrac{u(2-\delta)\Omega_{D}}{1-(2-\delta)\Omega_{D}}-1.
\end{eqnarray}
Now, bearing Eq.~(\ref{u}) in mind, this result takes finally the
form
\begin{eqnarray}\label{w1}
\omega_{D}=\dfrac{\delta-1}{(2-\delta)\Omega_{D}-1}.
\end{eqnarray}
For $\delta<1$, we have $2-\delta>1$ meaning that there is a divergence in the behavior of $\omega_{D}$ happen at the redshift for which $\Omega_{D}=\frac{1}{2-\delta}$. Therefore, the $\delta<1$ case is not suitable in our setup. From Eq.~(\ref{3}), we can easily find that
\begin{eqnarray}\label{8}
\Omega_{D}^{\prime}&&=\frac{d\Omega_{D}}{d(\ln a)}=(-2\delta+2)\Omega_{D}\frac{\dot{H}}{H^{2}}.
\end{eqnarray}
Combining this result with Eqs.~(\ref{80}) and~(\ref{w1}), one obtains
\begin{eqnarray}\label{9}
\Omega_{D}^{\prime}=3(\delta-1)\Omega_{D}\left(\dfrac{1-\Omega_{D}}{1-(2-\delta)\Omega_{D}}\right),
\end{eqnarray}
which finally leads to
\begin{eqnarray}
\Omega_{D}(1-\Omega_{D})^{1-\delta}=\mathcal{C}a^{3(\delta-1)}=\mathcal{C}(1+z)^{3(1-\delta)},
\end{eqnarray}
where $\mathcal{C}$ is the integration constant determined by initial conditions. We also used the $1+z=\frac{1}{a}$ relation, where $z$ denotes the redshift, in order to obtain the last equality.
\begin{figure}[ht]
\centering
\includegraphics[scale=0.35]{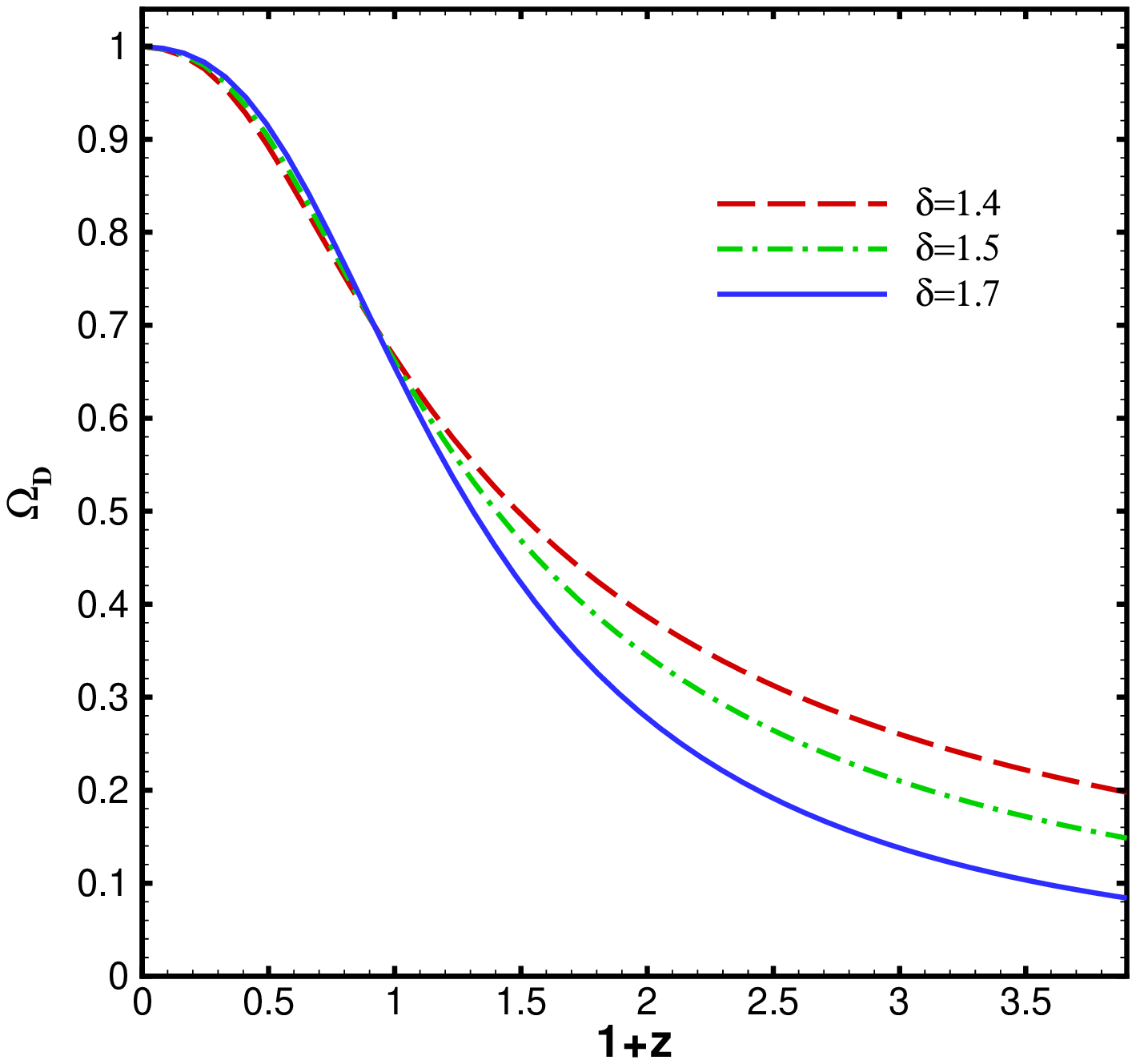}
\includegraphics[scale=0.35]{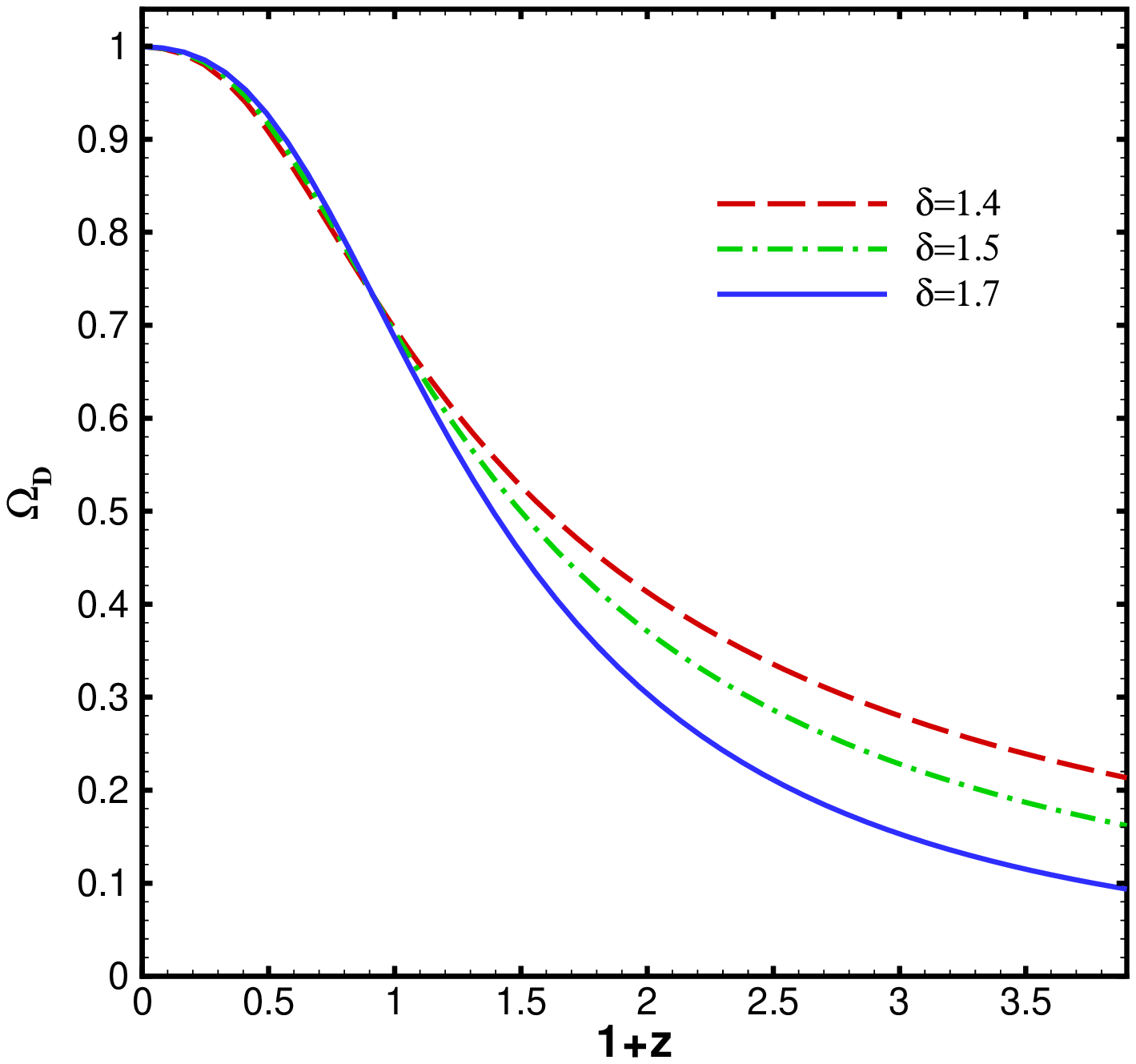}
\caption{$\Omega_D$ for some values of $\delta$. The
$\Omega_D(z=0)=0.70$ and $\Omega_D(z=0)=0.73$ conditions
have been used to plot the upper and lower panels, respectively.
\label{Omega}}
\end{figure}
Considering some values of $\delta$, the $\omega_D$ and $\Omega_D$
parameters have been plotted as the functions of the redshift $z$
in Figs.~\ref{w} and~\ref{Omega} for the $\Omega_D(z=0)=0.70$
(the upper panel) and $\Omega_D(z=0)=0.73$ (the lower panel) initial
conditions, where $z=0$(or equally a=1) represents the current era. It is easy to
see that for $\Omega_{D}\rightarrow0$ ($\Omega_{D}\rightarrow1$),
we have $\omega_{D}\rightarrow1-\delta$
($\omega_{D}\rightarrow-1$). Moreover, if $\delta=2$, then
$w_D=-1$, a result independent of the value of $\Omega_{D}$. In
fact, as it is apparent from Eqs.~(\ref{rho}) and~(\ref{6}), this
case is mathematically equivalent to the famous cosmological
constant model of DE.

We can also calculate the deceleration parameter, defined as
\begin{eqnarray}
q=-1-\frac{\dot{H}}{H^{2}},
\end{eqnarray}
by substituting Eq.~(\ref{w1}) into Eq.~(\ref{80}) as
\begin{figure}[ht]
\centering
\includegraphics[scale=0.35]{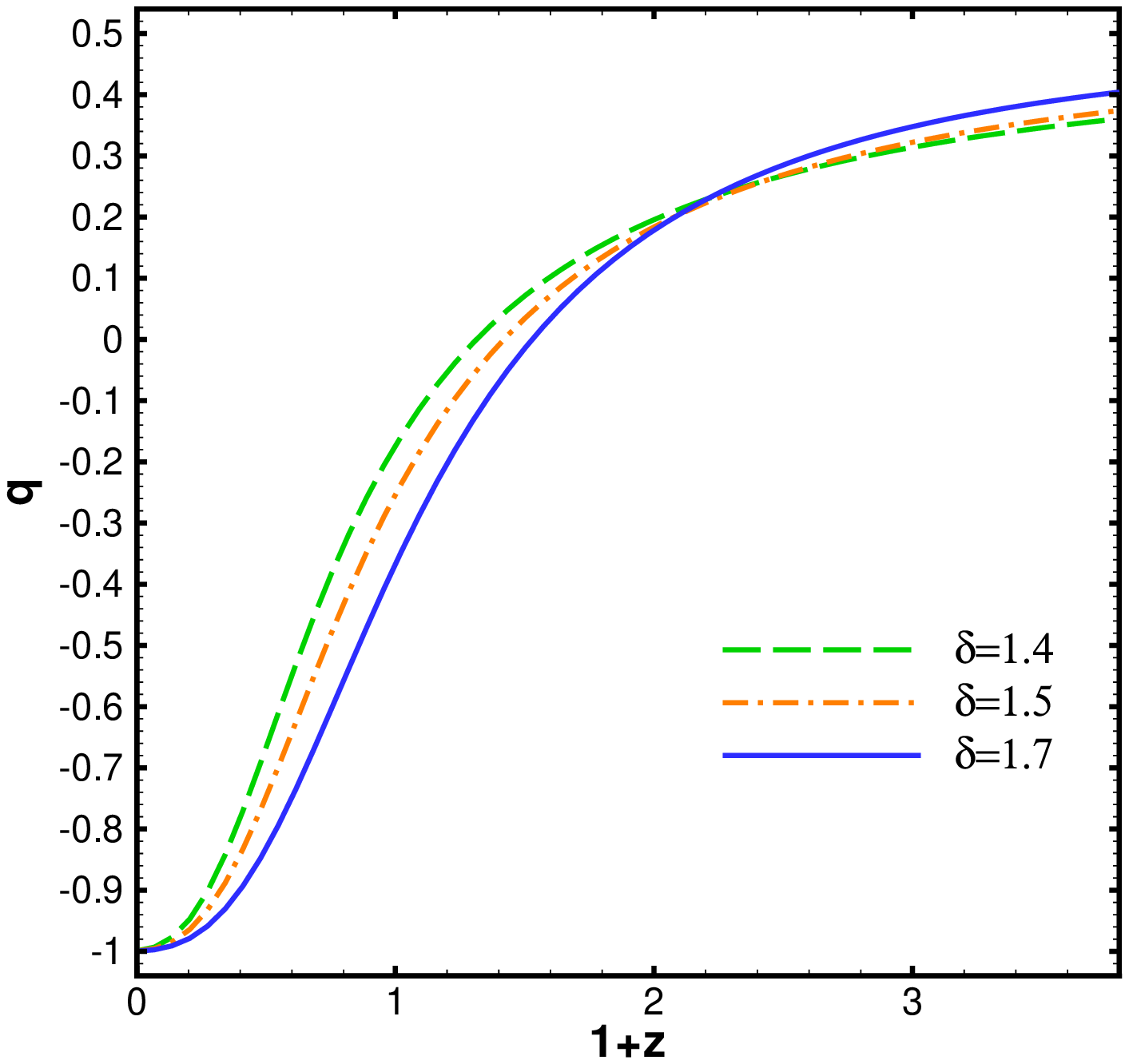}
\includegraphics[scale=0.35]{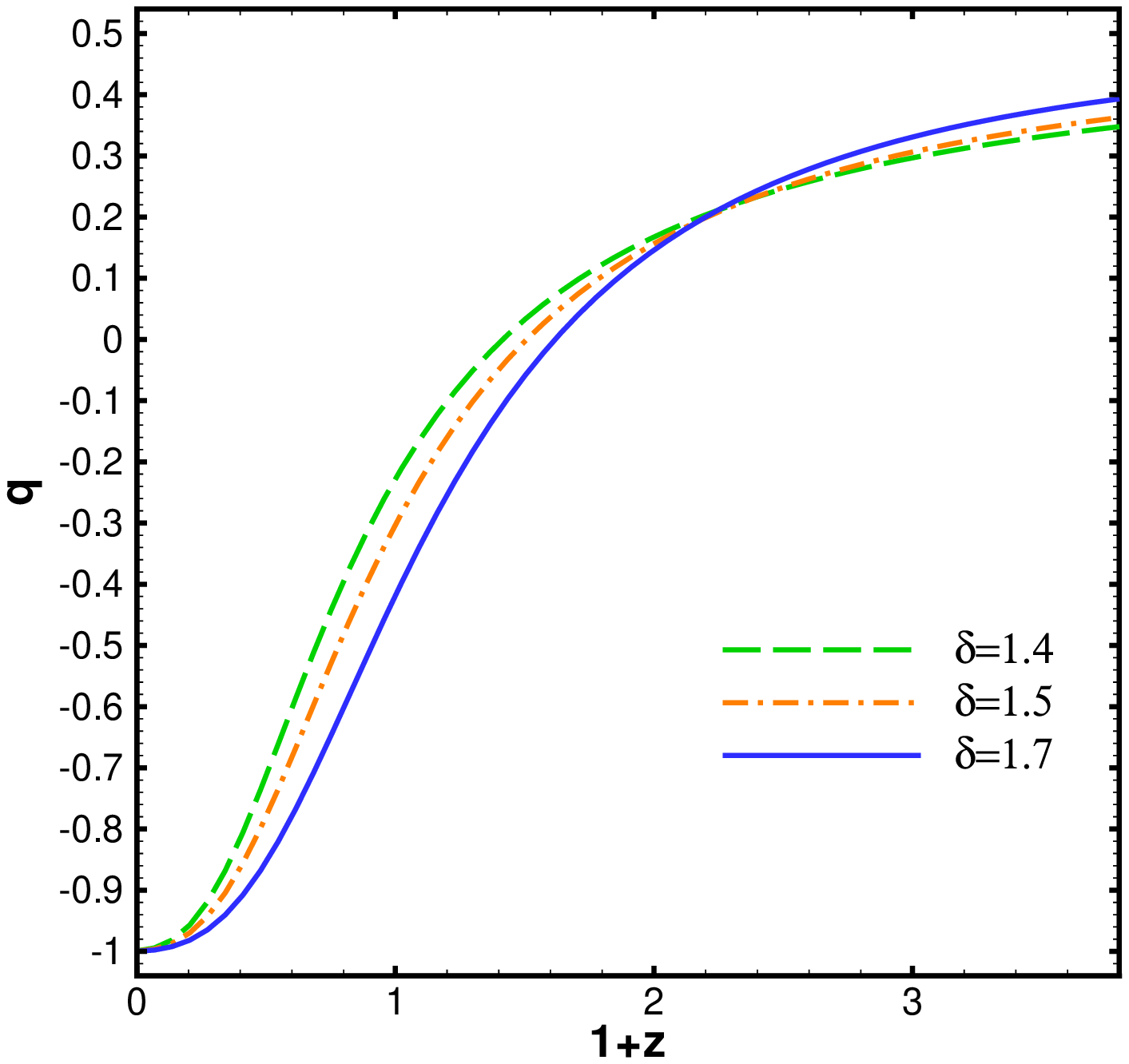}
\caption{$q$ for some values of $\delta$, when
$\Omega_D(z=0)=0.70$ (the upper panel) and
$\Omega_D(z=0)=0.73$ (the lower panel). \label{q}}
\end{figure}
\begin{eqnarray}\label{q}
q=\left(\frac{1}{2}\right)\left[\dfrac{(1-2\delta)\Omega_{D}+1}{1-(2-\delta)\Omega_{D}}\right].
\end{eqnarray}
The deceleration parameter $q$ has been plotted for some values of
$\delta$ in Fig.~\ref{q}, where $\Omega_D(z=0)=0.70$ (the
upper panel) and $\Omega_D(z=0)=0.73$ (the lower panel). It is
obvious that at $\Omega_{D}\rightarrow1$
($\Omega_{D}\rightarrow0$) limit, we have $q=-1$ ($q=1/2$), a
desired asymptotic behavior independent of the value of $\delta$.
It is also apparent that, depending on the value of
$\delta$, a suitable range for the transition redshift $z_t$ (from a decelerated to an
accelerated universe) is obtainable ($0.48\leq z_t<1$) \cite{Riess0,Riess,Riess1,Riess2,Riess3,COL2001,COL20011,COL20012,
COL20013,HAN2000,HAN20001,HAN20002}.

\begin{figure}[ht]
\centering
\includegraphics[scale=0.35]{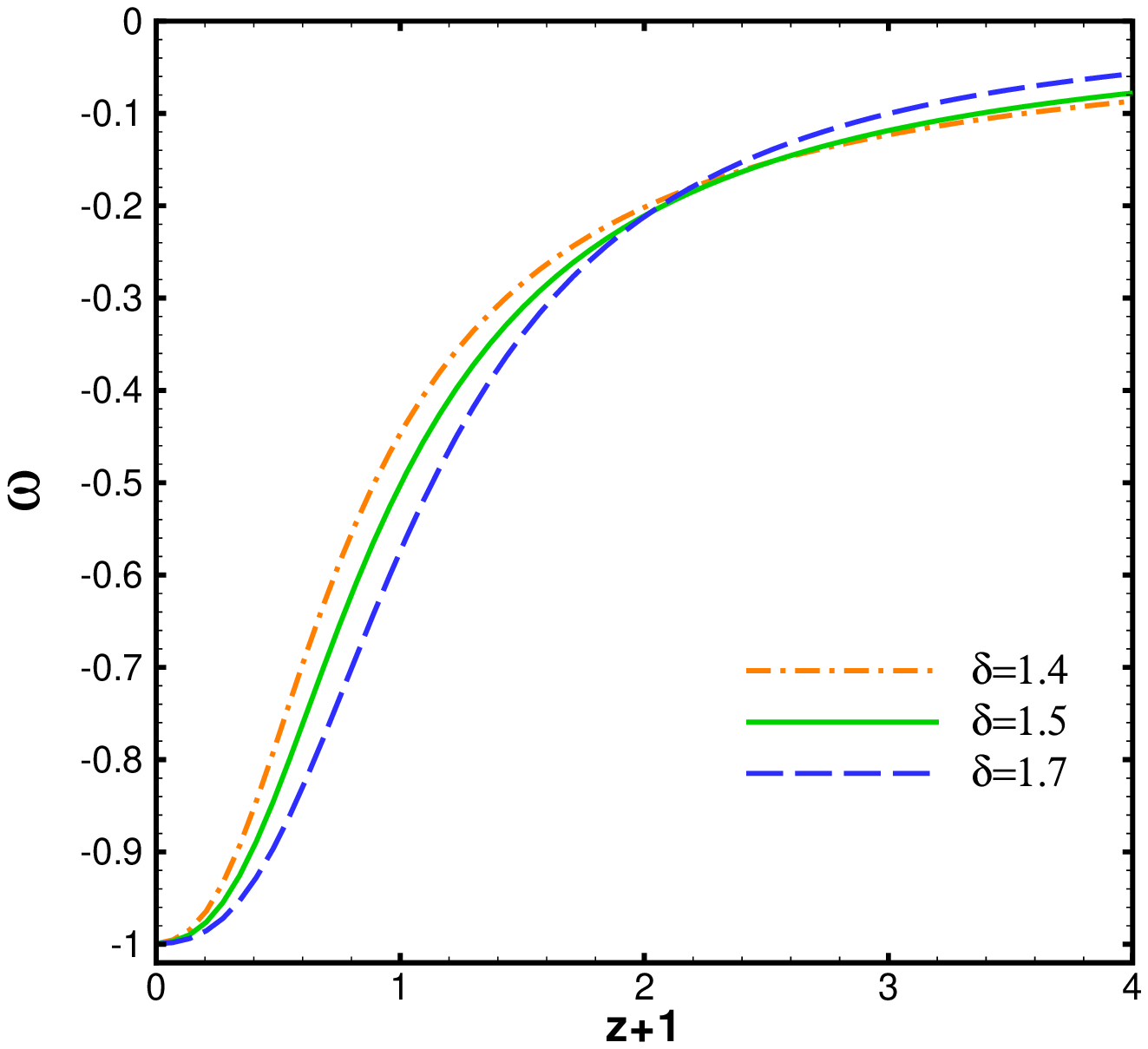}
\includegraphics[scale=0.35]{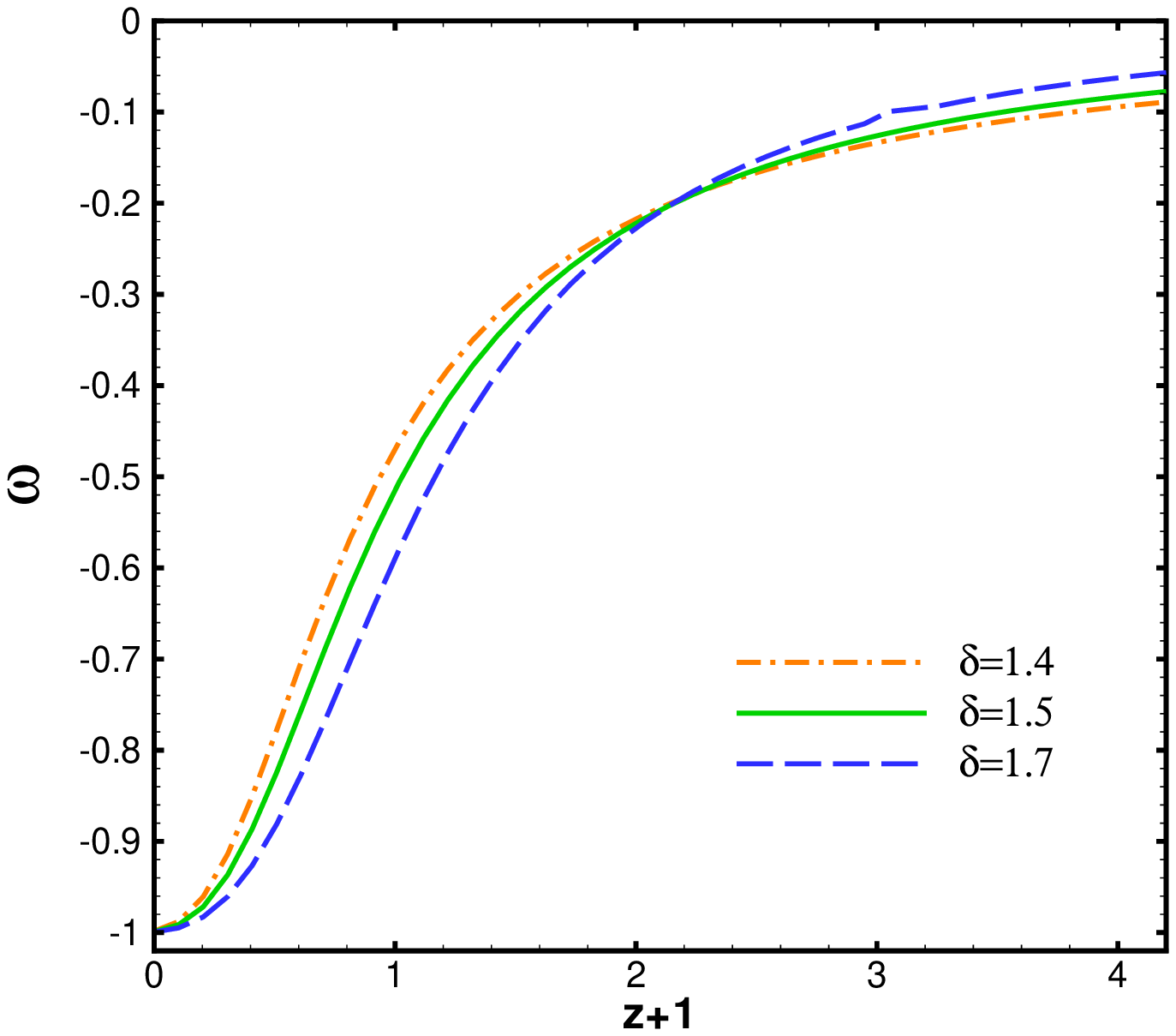}
\caption{$\omega$ for some values of $\delta$ as a function of
$z$, whenever $\Omega_D(z=0)=0.70$ (the upper panel) and
$\Omega_D(z=0)=0.73$ (the lower panel). \label{wt}}
\end{figure}

The total equation of state parameter is defined as
$\omega=P_{D}/(\rho_{D}+\rho_{m})$ leading to
\begin{eqnarray}
\omega=\left(\dfrac{1}{1+u}\right)\omega_{D}=\Omega_{D}\omega_{D},
\end{eqnarray}
plotted versus $z$ in Fig.~\ref{wt}. As a desired result, we see
that $\omega\rightarrow0$ ($\omega\rightarrow-1$) with increasing
(decreasing) $z$ in full agreement with this fact that the
pressureless DM was dominant at the early time in our model.
Finally, we explore the stability of the THDE model against
perturbation by considering its squared sound speed
$v_s^2=\frac{dP_D}{d\rho_D}$ whose sign determines the stability of
the model \cite{Peebles20031}. When $v_s^2<0$ the model is
unstable against perturbation. The squared sound speed is written
as
\begin{equation}\label{vs}
v_{s}^{2}=\frac{dP_D}{d\rho_D}=\frac{\dot{P}_D}{\dot{\rho}_D}=\dfrac{\rho_{D}}{\dot{\rho}_{D}}
\dot{\omega}_{D}+\omega_{D}.
\end{equation}
Now, using Eqs.~(\ref{rho}),~(\ref{8}) and~(\ref{w1}), one easily
finds
\begin{equation}\label{1}
\frac{\rho_{D}}{\dot{\rho}_{D}}=\frac{-1}{3H}\left(\dfrac{1-(2-\delta)
\Omega_{D}}{2-\delta-(2-\delta)\Omega_{D}}\right).
\end{equation}
Taking the time derivative of Eq.~(\ref{w1}), we reach
\begin{equation}\label{ne}
\dot{\omega}_{D}=\dfrac{(2-\delta)(1-\delta)H\Omega_{D}^{\prime}}{\left[1-(2-\delta)\Omega_{D}\right]^{2}},
\end{equation}
where we used the $\dot{\Omega}_{D}=H\Omega_{D}^{\prime}$ relation obtained from
Eq.~(\ref{8}). It can be combined with Eqs.~(\ref{9}),~(\ref{vs}) and~(\ref{1}) to reach at
\begin{eqnarray}
v_{s}^{2}=\dfrac{(\delta-1)(\Omega_{D}-1)}{\left[1-(2-\delta)\Omega_{D}\right]^{2}}.
\end{eqnarray}
Since $0<\Omega_D<1$, the sign of $v_{s}^{2}$ depends on the value of
$\delta$. For $\delta>1$, it is clear that $v_{s}^{2}<0$, and the
THDE is unstable, while for $\delta\leq1$ we have $v_{s}^{2}\geq0$
which implies that THDE with Hubble cutoff is stable against
perturbation. But, as it has been argued, the latter leads to a singularity in the behavior of $\omega_D$ in our setup.
\section{The Universe age}

The age of the present universe can be evaluated as
\begin{eqnarray}\label{0}
&&t=\int \frac{dtdH}{dH}=\int\frac{1}{\frac{\dot{H}}{H^2}}\frac{dH}{H^2}\\&&=\frac{(\frac{B}{3m_p^2})^{\frac{1}{2\delta-2}}}{3(1-\delta)}
\int\frac{(2-\delta)\Omega_D-1}{\Omega_D^{\frac{3-2\delta}{2-2\delta}}[1-\Omega_D]}d\Omega_D,\nonumber
\end{eqnarray}
where we used the $\frac{dH}{H^2}=\frac{(\frac{B}{3m_p^2})^{\frac{1}{2\delta-2}}d\Omega_D}{2(1-\delta)\Omega_D^{\frac{3-2\delta}{2-2\delta}}}$ relation as well as Eqs.~(12) and~(14) to obtain the above result. It finally leads to
\begin{eqnarray}
&&t=\frac{2(2-\delta)}{3H}\times\\ &&[1+\big(\frac{\delta-1}{2-\delta}\big)\ _2F_1(1,\frac{1}{2(\delta-1)};1+\frac{1}{2(\delta-1)};\frac{BH^{2-2\delta}}{3m_p^2})],\nonumber
\end{eqnarray}
in which $_2F_1(a,b;c;d)$ is the hypergeometric function of the $1$st kind. In order to have an estimation for the order of the age of the current universe ($z=0$), using the second equality of Eq.~(\ref{0}), and Eq.~(12), one may write
\begin{eqnarray}\label{00}
&&t=\int\frac{1}{\frac{\dot{H}}{H^2}}\frac{dH}{H^2}\approx\big(\frac{1}{\frac{\dot{H}}{H^2}}\big)\bigg|_{z=0}\int\frac{dH}{H^2}\\
&&=\frac{2(2-\delta)}{3H_0}\big(1-\frac{\omega_D(z=0)}{1+\omega_D(z=0)}\big)\nonumber,
\end{eqnarray}
where $H_0$ is the current value of the Hubble parameter. In this manner, if we consider the initial condition $\omega_D(z=0)=-2/3$, then we have $t=1/H_0$ for $\delta=3/2$. As another example, for $\Omega_D(z=0)=0.70$ and $\delta=1.7$, we have $\omega_D\simeq-0.87$ (from the upper panel of Fig.~\ref{w}) and thus $t\approx1.5/H_0$. A more accurate calculation by using Eq.~(\ref{0}) leads to $t\simeq0.92/H_0$ for this case meaning that Eq.~(\ref{00}) gives true order ($1/H_0$) for the universe age. Eq.~(\ref{00}) may also be modified by considering $i$) possible interactions between the cosmos sectors $ii$) other IR cutoffs $iii$) various corrections to the entropy or even $iv$) a combination of the mentioned ways.
\section{Conclusions}
Using the new entropy expression suggested by Tsallis and Cirto
\cite{non3}, and taking the HDE hypothesis into account
\cite{HDE,HDE5,HDE17}, we proposed a new holographic dark energy, namely THDE. In
addition, we considered a non-interacting flat FRW universe and
studied the evolution of the system. $q$, the total equation of state ($\omega$), the equation of
state of THDE ($\omega_D$), and $\Omega_D$ were studied. It has been found out
that this model can describe the late time acceleration of the
Universe expansion for some values of $\delta$($>1$). It should also be noted that, for $\delta>1$, the model is not stable at the classical level during the cosmic evolution (for all values of
$z$). This may be resolved by considering $i$) probable interactions between the
cosmos sectors $ii$) other IR cutoffs $iii$) various corrections
to the entropy or even $iv$) a combination of the these ways
\cite{RevH,wang,stab}. In fact, these considerations may also increase and modify the behavior and predictions (such as the present universe age) of THDE. They are subjects studied at the next steps to become more close to the various properties of THDE, and thus the origin of dark energy.
\acknowledgments{This work has been supported financially by
Research Institute for Astronomy \& Astrophysics of Maragha
(RIAAM). The work of Kazuharu Bamba is supported by
the JSPS KAKENHI Grant Number JP 25800136 and
Competitive Research Funds for Fukushima University
Faculty (17RI017).}

\end{document}